\documentclass[12pt]{article}
\begin{document}
\pagestyle{plain}
\title{Lorentz-Dirac equation in the delta-function pulse}
\author{Miroslav Pardy\\
Department of Physical Electronics \\
Masaryk University \\
Kotl\'{a}\v{r}sk\'{a} 2, 611 37 Brno, Czech Republic\\
e-mail:pamir@physics.muni.cz}
\date{\today}
\maketitle
\vspace{20mm}

\begin{abstract}
We formulate the Lorentz-Dirac equation in the plane wave and  in the Dirac $\delta$-function pulse. The discussion on the 
relation of the Dirac $\delta$-function to the ultrashort laser pulse is involved.
\end{abstract}

%%%%\newpage
\vspace{10mm}
\baselineskip 15 pt

\section{Introduction}

The problem of interaction an elementary particles with the laser field is,
at present time, one of the most prestige problem in the particle
physics. It is supposed that, in the future, the laser will
play the same role in particle physics as the linear or circular accelerators working in today particle laboratories.
The lasers nowadays provide one of the most powerful
sources of electromagnetic (EM) radiation under
laboratory conditions and thus inspire the fast growing
area of high field science aimed at the exploration of
novel physical processes (Mourou et al. 2006; Marklund et al., 2006; Salamin et al., 2006). 

Lasers have already demonstrated
the capability to generate light with the intensity
of ${\rm 2\times 10^{22}W/cm^{2}}$ (Yanovsky, et al., 2008) and projects to achieve ${\rm 10^{26} W/cm^{2}}$ (Dunne, 2006)
are under way. Further intensity growth towards and
above ${\rm 10^{23}W/cm^{2}}$ will bring us to experimentally unexplored
regimes. At such intensities the laser interaction
with matter becomes strongly dissipative, due to efficient
EM energy transformation into high energy gamma rays
(Zel'dovich, 1975; Zhidkov, et al. 2002). These gamma-photons in the laser field may produce
electron-positron pairs via the Breit-Wheeler process (Breit, et al. 1934). Then the pairs accelerated by the laser generate
high energy gamma quanta and so on (Bell, 2008), and thus
the conditions for the avalanche type discharge are produced
at the intensity ${\rm \approx  10^{25} W/cm^{2}}$. The occurrence
of such "showers" was foreseen by Heisenberg and Euler (1936).
A conclusion is made by Fedotov et al., (2010), that depletion of the laser energy on the electron-positron-gamma-ray plasma
(EPGP) creation could limit attainable EM wave intensity
and could prevent approaching the critical quantum
electrodynamics (QED) field. This field (Heisenberg et al., 1936; Sauter, 1931; Schwinger, 1951) is also
called the Schwinger field, $E_{S} = m^{2}c^{3}/e\hbar  = 1.32\times 10^{16} {\rm V/cm}$, corresponding to the intensity of ${\rm \approx 10^{29}W/cm^{2}}$.

One of the problem is acceleration of charged particles by the laser.
The acceleration effectiveness of the linear or circular accelerators is
limited not only by geometrical size of them
but also by the energy loss of accelerated particles
which is caused by bremsstrahlung during the
acceleration. The amount of radiation follows  from the Larmor formula for emission of radiation by accelerated charged particle 
(Landau et al.1988).

Here, we consider the interaction of an
electron with a plane wave field or, with a ultrashort (Dirac $\delta$-function) laser pulse. The quantum motion of electron in 
a plane wave was firstly described by Wolkow (1935). It must coincide in the classical limit with the classical solution.

First, we consider the classical approach to motion of a charged particle in a plane  field and then in a Dirac $\delta$-function  pulse. The attosecond  laser pulses  described 
in new time in article by Agostini et al.(2004) represents support
for activity in laser physics of the ultrashort laser pulses.

\section{Classical interaction of a charged particle with a plane wave}

To find motion of an electron in a periodic electromagnetic field, it is suitable to solve Lorentz equation
in general with four potential $A_{\mu} = a_{\mu}A(\varphi)$, where $\varphi = kx, k^{2} = 0$.
Following Meyer (1971) we apply his method and then we are prepared to consider radiation
reaction which has some influence on the motion of electron in the
electromagnetic field.

The Lorentz equation with $A_{\mu} = a_{\mu}A(\varphi)$ reads:

$$\frac {dp_{\mu}}{d\tau} = \frac {e}{m}F_{\mu\nu}p^{\nu}
= \frac {e}{m}(k_{\mu}a\cdot p - a_{\mu}k\cdot p)A'(\varphi); \quad A' = \frac{dA}{d\varphi}, 
\eqno(1)$$
where $\tau$ is proper time and $p_{\mu} = m(dx_{\mu}/d\tau)$.
After multiplication of the last equation by $k^{\mu}$, we get with regard to
the Lorentz condition $ 0 = \partial_{\mu}A^{\mu} =
a^{\mu}\partial_{\mu}A(\varphi) = k_{\mu}a^{\mu}A'$, or,
$k\cdot a = 0$ and $k^{2} = 0$, the following equation:

$$\frac {d(k\cdot p)}{d\tau} = 0\eqno(2)$$
and it means that $k\cdot p$ is a constant of the motion and it can be defined
by the initial conditions  for instance at time $\tau = 0$. If we put
$p_{\mu}(\tau = 0) = p_{\mu}^{0}$, then we can
write $k\cdot p = k\cdot p^{0}$. At this moment we have with $d\varphi = k\cdot dx$:

$$k\cdot p = \frac{mk\cdot dx}{d\tau} = m\frac {d \varphi}{d\tau},
\eqno(3)$$
or,

$$\frac{d\varphi}{d\tau} = \frac {k\cdot p^{0}}{m}.
\eqno(4)$$

So, using the last equation and relation $d/d\tau =
(d/d\varphi)d\varphi/d\tau$, we can write equation (1) in the form
$(dp_{\mu}/d\tau  =(d\varphi/d\tau)(d p_{\mu}/d\varphi))$:

$$\frac {dp_{\mu}}{d\varphi} = \frac {e}{k\cdot p^{0}}
(k_{\mu}a\cdot p - a_{\mu}k\cdot p^{0})A'(\varphi) = 
e\left(k_{\mu}\frac{a\cdotp}{k\cdotp^{0}} - a_{\mu}\right)A'(\varphi)\eqno(5)$$
giving (after multiplication by $a^{\mu}$)

$$\frac {d(a\cdot p)}{d\varphi} = - ea^{2}A' ,\eqno(6)$$
or,

$$a\cdot p = a\cdot p^{0} - ea^{2}A.\eqno(7)$$

Substituting the last formula into (5), we get:

$$\frac {dp_{\mu}}{d\varphi} = -e\left(a_{\mu} -
\frac {k_{\mu}a\cdot p^{0}}{k\cdot
p^{0}}\right)\frac {dA}{d\varphi} -
\frac {e^{2}a^{2}}{2k\cdot p^{0}}\frac {d(A^{2})}{d\varphi}
k_{\mu}.
\eqno(8)$$

This equation  can be immediately integrated to
give the resulting momentum in the
form:

$$p_{\mu} = p_{\mu}^{0} - e\left(
A_{\mu} - \frac {A^{\nu}p_{\nu}^{0}k_{\mu}}{k\cdot p^{0}}\right)
- \frac {e^{2}A^{\nu}A_{\nu}k_{\mu}}{2k\cdot p^{0}}.\eqno(9)$$

\section{Classical interaction of a charged particle with a $\delta$-function pulse}

One of the primary goals of ultrashort laser science is to provide more insights into the dynamics of atomic electrons. One general interest is the direct probing in time of hyperfast electronic rearrangements following the creation of an inner-shell
hole. There is a  study using sub-femtosecond burst of XUV light
probed the motion of an electron wave packet under the influence of
an infrared laser's electric field. Furthermore, the precise timing of the electron wave packet emitting the high harmonics can be measured by observing the two-photon ionization electron
energy spectrum These pioneering experiments are
reviewed  by Agostini et al. (2004). 

What are the limits and future of attosecond pulses? 
The goal of the laser physics is to generate very short pulses. At
present time we are able to generate the  attosecond pulses (Agostini et al., 2004). Nevertheless, the final goal of short pulse laser laboratories is to generate laser pulse in the  $\delta$-function form, and there is no theory which restricts the attainability of such pulses. It is not excluded that the mystery of the Higgs boson will be revealed just using such laser pulses. So, let us first remember the rigorous theory of $\delta$-function.

The $\delta$-function mathematical theory can be presented in a very simple way (Martynenko, 1973).  Namely, using the definition of the unite Heaviside step function denoted as the $\eta$-function. It is defined by
the relation:

$$\eta(t) = \left\{ \begin{array}{c} 0, \quad t <  0  \\
1, \quad t \geq 0
\end{array} \right. .
\eqno(10)$$

The $\eta$-function is the limiting case of the sequence  $\eta_{n}(t)$.

$$\eta_{n}(t) = \frac{1}{2} + \frac{1}{\pi}\arctan(nt); \quad
|\arctan(nt)| < \frac{\pi}{2}. \eqno(11)$$

Using this sequence, we define the $\delta$-generating sequence 

$$\delta_{n}(t) = \frac{d}{dt}\eta_{n}(t),\eqno(12)$$ 
or, 

$$\delta_{n}(t) = \frac{n}{\pi(t^{2}n^{2} + 1)}. \eqno(13)$$ 

The $\delta$-function is then defined as the limiting case of the last relation 

$$\delta(t) = \lim_{n\rightarrow \infty} \frac{n}{\pi(t^{2}n^{2} + 1)}.\eqno(14)$$ 

So, we get that the $\delta$-function is derived as

$$\delta(t) = \left\{ \begin{array}{c} 0, \quad t \neq  0  \\
\infty , \quad t = 0
\end{array} \right..
\eqno(15)$$ 

If we perform the integration of the $\delta_{n}$ function, we get:

$$\int_{-\infty}^{\infty}\delta_{n}(t)dt = \int_{-\infty}^{\infty}\frac{n}{\pi(t^{2}n^{2} + 1)}dt = \frac{2}{\pi}\int_{0}^{\infty}d(\arctan(nt)) = 1 .\eqno(16)$$

For $t = 0$, we have

$$\delta_{n}(t) = \frac{n}{\pi}; \quad \delta(t) = 
\lim_{n\rightarrow \infty}\delta_{n}(t)\eqno(17)$$
and 

$$ \int_{-\infty}^{\infty}\delta(t)dt =  1.  \eqno(18)$$

So, we write

$$\frac{d}{dt}\eta(t) = \delta(t) = \left\{ \begin{array}{c} 0, \quad t \neq  0  \\
\infty , \quad t = 0
\end{array} \right.;\quad \int_{-\infty}^{\infty}\delta(t)dt = 1.
\eqno(19)$$ 

Let us remark that the $\delta$-function was in the history of
mathematics used also by Poisson, Cauchy, Hermite and others. At
present time the $\delta$-function is called the Dirac $\delta$-function
because it was introduced into quantum mechanics rigorously by Dirac.

The $\delta$-function has also meaning in classical mechanics.
Newton's second law
for the interaction of a massive particle with mass $m$ with an impact force $\delta(t)$
is as follows:

$$m\frac{d^{2}x}{dt^{2}} =  P\delta(t),\eqno (20)$$
where P is some constant. If we express $\delta$-function by the relation $\delta(t) =\dot\eta(t)$,
then from eq. (20) $\dot x(t)= P/m$ follows immediately. The physical meaning of the quantity $P$ can be deduced from equation $F = P\delta(t)$. After $t$-integration we have $\int F dt  = \int m(dv/dt)dt = mv =  P$, where $m$ is mass of a body and $v$ its final velocity (with $v(0)= 0$). It means that the value of $P$ can be determined a posteriori and then this value can be used in more complex equations than eq. (20). Of course it is
necessary to suppose that $\delta$-form of the impact force is adequate approximation of the experimental situation.

Now, if we put into formula (9) the four-potential $A_{\mu} = a_{\mu}A(\varphi) = a_{\mu}\eta(\varphi)$
of the impact force,
then for $\varphi\ > 0$ when $\eta > 1$, we get:

$$p_{\mu} = p_{\mu}^{0} - e\left(
a_{\mu} - \frac {a^{\nu}p_{\nu}^{0}k_{\mu}}{k\cdot p^{0}}\right)
- \frac {e^{2}a^{\nu}a_{\nu}k_{\mu}}{2k\cdot p^{0}}.
\eqno(21)$$

The last equation can be used to determination of the magnitude of $a_{\mu}$
similarly as it was done in discussion to the eq. (20). It can be evidently expressed as the number of $k$-photons in electromagnetic momentum. For
$\varphi < 0$, it is $\eta = 0$ and therefore $p_{\mu} = p_{\mu}^{0}$

It is still necessary to say what is the practical realization of the
$\delta$-form potential. We know from the Fourier analysis
that the Dirac $\delta$-function can be expressed by integral
in the following form:

$$\delta(\varphi) = \frac {1}{\pi}\int_{0}^{\infty}\cos(s\varphi)ds.
\eqno(22)$$

So, the $\delta$-force and $\delta$-potential can be realized as the continual
superposition of the harmonic waves. In case it will be not possible
to realize  experimentally it,
we can approximate the integral formula by the summation
formula as follows:

$$\delta(\varphi) \approx \frac {1}{\pi}\sum_{0}^{\infty}\cos(s\varphi).
\eqno(23)$$

If we consider the $\delta$-form electromagnetic pulse, then we can write

$$F_{\mu\nu} = a_{\mu\nu}\delta(\varphi).\eqno(24)$$
where $\varphi = kx = \omega t - {\bf k}{\bf x}$. In order to obtain the
electromagnetic impulsive force in this form, it is necessary
to define the four-potential in the following form:

$$A_{\mu} = a_{\mu}\eta(\varphi),\eqno(25)$$
where function $\eta$ is the Heaviside unit step  function defined by
the relation:

$$\eta(\varphi) = \left\{ \begin{array}{c} 0, \quad \varphi <  0  \\
1, \quad \varphi \geq 0
\end{array} \right. .
\eqno(26)$$

If we define the four-potential by the equation (25), then the
electromagnetic tensor with impulsive force is of the form:

$$F_{\mu\nu} = \partial_{\mu}A_{\nu} - \partial_{\nu}A_{\mu} =
(k_{\mu}a_{\nu} - k_{\nu}a_{\mu})\delta(\varphi) =
a_{\mu\nu}\delta(\varphi).\eqno(27)$$

\section{Lorentz-Dirac equation in a plane wave and in the $\delta$-function field}

It is well known that Lorentz-Dirac equation describes motion of a charged particle in electromagnetic field where also the bremsstrahlung force is involved in the equation.

The bremsstrahlung force is in the nonrelativistic limit expressed in the following form:

$${\bf f} = \frac {2e^{2}}{3c^{3}}\ddot{\bf v}.\eqno(28)$$

The force ${\bf f}$ is not active force and it means it cannot cause motion
of electron. In other words the equation ${\bf f} = m\dot{\bf v}$ has no
physical meaning. The bremsstrahlung force is meaningful only with
addition of the active force. Then the bremsstrahlung force acts as the so called light friction.

The relativistic equation which involves the bremsstrahlung force is so called
Lorentz-Dirac equation and it can  be evidently written in the form
(Landau et al., 1988):

$$mc\frac {dv_{\mu}}{d\tau} = \frac {e}{c}F_{\mu\nu}v^{\nu}
+ g_{\mu},\eqno(29)$$
where $g^{\mu}$ can expressed according to Landau et al. (1988) in the form:

$$g_{\mu} = \frac {2e^{2}}{3c}\left(\frac {d^{2}v_{\mu}}{d\tau^{2}}
 - v_{\mu}v^{\nu}\frac {d^{2}v_{\nu}}{d\tau^{2}}\right),\eqno(30)$$
where the form of the bremsstrahlung term leads in the nonrelativistic limit to eq. (28),
where $u_{\mu}$ is the four-velocity and the radiative term was approximated by Landau et al. in the form (Landau et al., 1988):

$$g_{\mu} = \frac{2e^{3}}{3mc^{3}}\frac{\partial F_{\mu\nu}}
{\partial x^{\alpha}}v^{\nu}v^{\alpha} - 
\frac{2e^{4}}{3m^{2}c^{5}} F_{\mu\alpha}F^{\beta\alpha}v_{\beta} + 
\frac{2e^{4}}{3m^{2}c^{5}} \left(F_{\alpha\beta}v^{\beta}\right) 
\left(F^{\alpha\gamma}v_{\gamma}\right)v_{\mu}. \eqno(31)$$

It is possible to show that the space components of the 4-vector force $g_{\mu}$ is of the form (Landau et al., 1988) 

$${\bf f} = \frac{2e^{3}}{3mc^{3}}\left(1 - \frac{v^{2}}{c^{2}}\right)^{-1/2}\left\{\left(\frac{\partial}{\partial t} + ({\bf v}\nabla)\right){\bf E}
+ \frac{1}{c}\left[{\bf v}\left(\frac{\partial}{\partial t} + ({\bf v}\nabla)\right){\bf H}\right]\right\} + $$

$$+ \frac{2e^{4}}{3m^{2}c^{4}}\left\{{\bf E}\times {\bf H} + \frac{1}{c}{\bf H }\times ({\bf H}\times {\bf v}) + \frac{1}{c}{\bf E}({\bf v}{\bf E})\right\} - $$

$$- \frac{2e^{4}}{3m^{2}c^{5}\left(1 - \frac{v^{2}}{c^{2}}\right)}{\bf v}\left\{\left({\bf E} + \frac{1}{c}({\bf v}\times{\bf H})\right)^{2} -
\frac{1}{c^{2}}({\bf E}{\bf v})^{2}\right\}.\eqno(32)$$

Using the Lorentz equation we can express the second derivative in the form:

$$\frac {d^{2}v_{\mu}}{d\tau^{2}} = \frac {e}{mc^{2}}\partial_{\alpha}
\left(F_{\mu\nu}\right)v^{\alpha}v^{\nu} +  \frac {e^{2}}{mc^{4}}F_{\mu\nu}F^{\nu\alpha}
v_{\alpha}.\eqno(33)$$

After insertion of the last equation in the Lorentz-Dirac equation, we get the Lorentz-Dirac equation in the final form:

$$mc\frac {dv_{\mu}}{d\tau} = \frac {e}{c}F_{\mu\nu}v^{\nu}
+ \frac {2e^{3}}{3mc^{3}}\left\{\partial_{\alpha}F_{\mu\nu}v^{\nu}v^{\alpha}
- \frac {e}{mc^{2}}F_{\mu\alpha}F^{\nu\alpha}v_{\nu} + \frac {e}{mc^{2}}
v_{\mu}F_{\alpha\beta}v^{\beta}F^{\alpha\gamma}v_{\gamma}\right\},\eqno(34)$$

It seems that the solution for the plane wave 
of the last equation was never given in
literature, so we write here such equation in order to find the solution by the same method which was used for the solution of the Lorentz equation.

Using plane wave potential $A_{\mu} = a_{\mu}A(\varphi)$ with $\varphi = kx$,
we get after insertion it into the last equation as follows:

$$mc\frac {dv_{\mu}}{d\tau} = \frac {e}{c}(k_{\mu}a_{\nu} - k_{\nu}a_{\mu})
A'v^{\nu} + \frac {2e^{3}}{3mc^{3}}\left\{(k_{\mu}a_{\nu} - k_{\nu}a_{\mu})
k_{\alpha}A''v^{\alpha}v^{\nu} - \right.$$

$$- \frac {e}{mc^{2}}(k_{\mu}a_{\alpha} - k_{\alpha}a_{\mu})
(k^{\nu}a^{\alpha} - k^{\alpha}a^{\nu})v_{\nu}A'^{2} + $$

$$+ \left. (k_{\alpha}a_{\beta} - k_{\beta}a_{\alpha})
(k^{\alpha}a^{\gamma} - k^{\gamma}a^{\alpha})
v^{\beta}v_{\gamma}v_{\mu}A'^{2}\right\}. \eqno(35)$$

If we define new quantities $K_{\mu\nu}$ as follows, $K_{\mu\nu} =
k_{\mu}a_{\nu} - k_{\nu}a_{\mu}$, we can write the last equation in the form:

$$mc\frac {dv_{\mu}}{d\tau} = \frac {e}{c}K_{\mu\nu}
A'v^{\nu} + \frac {2e^{3}}{3mc^{3}}\left\{K_{\mu\nu}
k_{\alpha}A''v^{\alpha}v^{\nu} - \right.$$

$$- \left.\frac {e}{mc^{2}}K_{\mu\alpha}K^{\nu\alpha}v_{\nu}A'^{2} + 
K_{\alpha\beta}K^{\alpha\gamma}
v^{\beta}v_{\gamma}v_{\mu}A'^{2}\right\}, \eqno(36)$$
which is the system of the differential equation which is interesting to be solved in mathematical physics.

In case of the periodic (laser field) we define the vector potential
in the form with polarization

$$A_{\mu}(\varphi) = a_{\mu}\sin\varphi + b_{\mu}\cos\varphi; \quad
\varphi = kx, \eqno(37)$$
where  for circularly polarized wave, the coefficients in eq. (37) are not independent but are defined by the relations:

$$a_{\mu}^{2} = b_{\mu}^{2}; \quad a_{\mu}b_{\mu} = 0.\eqno(38)$$

For the $\delta$-function pulse it is $A_{\mu} = a_{\mu}\eta(\varphi)$. 

In case that we are interested to determine function $F = kv$ from the differential equation, we multiply eq. (35) by $k^{\mu}$, and  we get with regard to relations $k^{2} = 0, ka = 0$:

$$mc\frac {dF}{d\tau} = \left(\frac{2e^{3}}{3mc^{3}}\right)\left(\frac
{e}{mc^{2}}\right)F^{3}a^{2}A'^{2},\eqno(39)$$
where $F = kv$.

After multiplication of equation (35) by $a^{\mu}$, we get with $G = av$:

$$mc\frac {dG}{d\tau} =  - \frac {e}{c}Fa^{2}A' +
\left(\frac{2e^{3}}{3mc^{3}}\right)\left\{-F^{2}a^{2}A'' +
\frac {e}{mc^{2}}F^{2}Ga^{2}A'^{2}\right\}\eqno(40)$$

In case of the $\delta$-function potential, we use the tensor of he electromagnetic field
$F_{\mu\nu}$ in the form (27) and the corresponding equations  for $F, G$ are as follows. 

$$mc\frac {dF}{d\tau} = \left(\frac{2e^{3}}{3mc^{3}}\right)\left(\frac
{e}{mc^{2}}\right)F^{2}a^{2}\eta'^{2}; \quad F = kv,\eqno(41)$$
where $F = kv$ and 

$$mc\frac {dG}{d\tau} =  - \frac {e}{c}Fa^{2}\eta' +
\left(\frac{2e^{3}}{3mc^{3}}\right)\left\{-F^{2}a^{2}\eta'' +
\frac {e}{mc^{2}}F^{2}Ga^{2}\eta'^{2}\right\};
\quad G = av,\eqno(42)$$
where for function $\eta$ we can use the definition (11) with $n\rightarrow \infty$, or directly the $\delta$-function for $\eta'$.

\section{Discussion}

We have presented, in this article,
the solution of the Lorentz equation for motion of a charged
particle in the electromagnetic plane wave field and in the $\delta$-function form of the laser pulse. We also formulated Lorentz-Dirac equation in the electromagnetic plane wave field and in the $\delta$-function form of the laser pulse.

The present article can be also related to the author
discussion on electron in laser field
(Pardy, 1998; Pardy, 2001), where the Compton model of laser acceleration was proposed and quantum theory of motion of electron in laser field.

The $\delta$-form laser pulses are here considered as an idealization of the experimental situation in laser physics. Nevertheless, it was
demonstrated theoretically that at present time the zeptosecond and
subzeptosecond laser pulses of duration $10^{-21} - 10^{-22}$ s can be realized by the petawat lasers. It means that the generation of the ultrashort laser pulses is the keen interest in development of laser physics (Agostini et al., 2004).

Let us remark that while the $\delta$-form pulses are not still used in the theoretical laser physics, such exotic pulses are 
constantly used in the synchrotron physics, where the equation for the betatron radial vibration involves the derivative of the $\delta$ function:

$$r'' + \frac{c^{2}}{R^{2}}r = \frac{1}{E}\sum_{i}\hbar cn
\delta'(t-t_{i}), \eqno(43)$$
where $r$ being the radial deflection from radius $R$ of the local orbit with energy $E$ and the  number of harmonic  $n$.

So the synchrotron theory uses not only $\delta$-form pulses of photons radiated by an electron accelerated on an orbit, but also their derivative (here denoted by symbol $'$).

The friction force in eq. (28) is not active force and it means it cannot cause motion of electron. In other words the equation ${\bf f} = m\dot{\bf v}$ has no physical meaning for ${\bf f}$ defined by eq. (28). The bremsstrahlung  force equation  is meaningful only with addition of the active force. Then, the bremsstrahlung force acts as the so called light friction.

The equation with light friction and absence of external force  

$$m\dot{\bf v} = \frac {2e^{2}}{3c^{3}}\ddot{\bf v}\eqno(44)$$ 
produces not only physical meaningful solution $\dot{\bf v} = const$ but also so called run-away solutions $\dot{\bf v} = \exp[(2e^{2}/3c^{3})t]$, the absurdity of  which is evident. The problem is frequently discussed in the scientific physical journals and it was not till this time solved by the satisfactory way. There is an elementary solution of this problem if we redefine the  friction force by relation 

$${\bf f} \rightarrow {\bf f}\eta(E); \quad E = |\bf E|.\eqno(45)$$

Let us remark that such elementary operation is not involved in the textbooks on electrodynamics. On the other hand, the adequate solution of the problems with Lorentz-Dirac equation is evidently in replacing this equation by quantum electrodynamics (QED) equation. It follows from the well known historical analogue. Namely, that he Bohr
planetary model equation for the hydrogen atom was replaced by the Schr\"odinger equation,
Schr\"odinger equation was replaced by the Pauli equation, Pauli equation was replaced by the Dirac equation and by the Dirac-Schwinger equation and QED, and QED was replaced by the Schwinger source theory including also gravity. So the final goal should be to find the QED-Lorentz-Dirac equation.

In case of an acceleration of a charged particle by the homogeneous 
gravitational field, no electromagnetic field is generated and there is no radiation reaction. Particle during its free fall is "free" and its surrounding field has energy, which corresponds to mass which falls as a free mass. The same conclusion should follow from the electro-gravity Einstein equations.   

We have seen that the $\delta$-function form of force is an impact which causes that the body obtains the nonzero velocity or nonzero momentum at $t=0$. The situation in quantum field theory is a such that  $\delta$-function is a source which can generate elementary particles. It is not excluded that the Big Bang started at $t=0$ by  $\delta$-function form of impact. The idea that the existence of  universe started with the zero radius was formulated many years ago by Friedmann and Lama$\hat{\rm i}$tre. 
While the Friedmann
solution follows from the Einstein general relativity, quantum chromodynamics gives  no answer that the Big Bang started by the   
$\delta$-function form  source of quarks and leptons.

In QED the charged particle interaction with EM fields
is determined by relativistically and gauge invariant function parameters $\chi, f, g$, and the total probability of emission of photons is function  $W(\chi, f, g)$, where (Ritus, 1979; Nikishov et al. 1970; Berestetzkii et al. 1089) ($\hbar = c = 1$)
 
$$\chi^{2} = -\frac{e^{2}}{m^{6}}(F_{\mu\nu}p^{\nu})^{2}, 
\quad f = -\frac{e^{2}}{m^{4}}(F_{\mu\nu})^{2}, \quad 
g = -\frac{e^{2}}{m^{4}}\varepsilon_{\lambda\mu\nu\varrho} 
F^{\lambda\mu}F^{\nu\varrho}.\eqno(46)$$

In case of the Redmond configuration, i.e.  in case of crossed fields, ${\bf E} \perp {\bf H}$, parameters $f, g =0$, and $W = W(\chi, 0, 0)$. This function is also the probability of the same process in the plane wave field (Nikishov et al., 1967). And,  It is possible to show that every ultrarelativistic electron moves in the situation with the fields of the Redmond configuration (Berestetzkii et al., 1989).

With regard to the fact that every invariant parameter involves the electromagnetic tensor $F^{\mu\nu}$, it can be expressed as the electromagnetic tensor for the $\delta$-function pulse, and also the invariant function $W$ can be transformed to this situation. With regard to eq. (27), we have ($\hbar = c = 1$)

$$\chi^{2} = -\frac{e^{2}}{m^{6}}(a_{\mu\nu}p^{\nu})^{2} \delta^{2}(\varphi), 
\quad f = -\frac{e^{2}}{m^{4}}(a_{\mu\nu})^{2}\delta^{2}(\varphi), \quad 
g = -\frac{e^{2}}{m^{4}}\varepsilon_{\lambda\mu\nu\varrho} 
a^{\lambda\mu}a^{\nu\varrho}\delta^{2}(\varphi).\eqno(47)$$

For the Redmond configuration we then have $W = W(\chi, 0, 0) \rightarrow F(\delta(\varphi))$. Such mathematical object is still not the integral part of the infinitesimal calculus and it means that the mathematical analysis is open for study of such objects. On the other hand, the adequate laser and particle physics corresponding to such object can be promising. 
 
New experiments can be realized and new measurements performed by means of
the ultrashort laser pulses, giving new results and discoveries.
So, it is obvious that the interaction of particles with the laser pulses can form, in the near future, the integral part of the laser and particle physics in such laboratories as ESRF, CERN, DESY, SLAC and specially ELI.

\vspace{10mm}

\noindent
{\bf REFERENCES}

\vspace{7mm}

\noindent
Agostini, P. and DiMauro, L. F. (2004). The physics of attosecond light pulses, Rep. Prog. Phys. {\bf 67}, 813–855. \\[2mm]
Berestetzkii, V. B, Lifshitz, E. M. and  Pitaevskii, L. P. (1989). \\ {\it Quantum Electrodynamics}, (Moscow, Nauka). \\[2mm]
Bell, A. R. and Kirk, J. G. (2008). Possibility of prolific pair production with high-power lasers, Phys. Rev. Lett. {\bf 101}, 200403.  \\[2mm]
Breit,  G. and  Wheeler, J. A. (1934). Collision of two light quanta, Phys. Rev. {\bf 46}, 1087–1091.  \\[2mm]
Dunne,  M. A. (2006). High-power laser fusion facility for Europe,
Nature Phys. {\bf 2}, 2-5. \\[2mm]
Fedotov,  A. M., Narozhny, N. B., Mourou, G. and Korn, G. (2010). Limitations on the attainable intensity of high power lasers,  Phys. Rev. Lett. {\bf 105}, 080402. \\[2mm]
Heisenberg,  W. and  Euler, H. Z. (1936). Folgerungen aus der
Diracschen Theorie des Positrons, Z. Phys. {\bf 98}, Nos. 11-12, 714-732. \\[2mm]
Landau, L. D. and  Lifshitz, E. M. (1988). {\it The Classical \\
Theory of Fields}, 4-th ed.~(London, Oxford, 1988).\\[2mm]
Marklund, M. and  Shukla, P. K. (2006). Nonlinear collective effects in photon-photon and photon-plasma interactions, Rev. Mod. Phys. {\bf 78}, 591.  \\[2mm]
Martynenko, V. S. (1973). {\it Operator Calculus}, (Kiev, 1973). (in Russian).  \\[2mm]
Meyer, J. W. (1971). Covariant classical motion of electron in a laser beam, Phys. Rev. D {\bf 3}, No. 2, 621 - 622. \\[2mm]
Mourou, G. A., Tajima, T. and  Bulanov, S. V. (2006). Optics in the relativistic regime Rev. Mod. Phys. {\bf 78}, Issue 2, 309. 
\\[2mm]
Nikishov, A. I. and  Ritus, V. I. (1967). Pair production by a photon and photon emission by an electron in the field of  an intense  electromagnetic wave and in a constant field, Soviet Phys. JETP {\bf 25}, No. 6, 1135.\\[2mm]
Nikishov, A. I. and  Ritus, V. I. (1970).  Quantum electrodynamics of strong fields, Sov. Phys. Usp. {\bf 13}, 303. \\[2mm]
Pardy, M. (1998). The quantum field theory of laser acceleration,
Phys. Lett. A {\bf 243}, 223-228. \\[2mm]
Pardy, M. (2001). The quantum electrodynamics of laser acceleration,
Radiation Physics and Chemistry {\bf 61}, 391-394.\\[2mm]
Reiss, H. R. (1962). Absorption of light by light, J. Math. Phys. {\bf 3}, 59. \\[2mm]
Ritus, V. I. (1979). The quantum effects of the interaction of elementary particles with the intense electromagnetic field, Trudy FIAN {\bf 111}, 5-151. (in Russian).  \\[2mm]
Salamin, Y. I.,  Hu, S.X.,  Hatsagortsyan, K. Z. and  Keitel, C. H. (2006). Relativistic high-power laser–matter interactions, Phys. Rep. {\bf 427}, Issues 2–3, 41-155. \\[2mm]
Sauter, F. (1931). $\ddot{\rm U}$ber Das Verhalten eines Electrons im homogenen elektrischen Feld nach der relativischen Theorie Diracs, (On the behavior of an electron in a homogeneous
electrical field according to the relativistic Theory
of Dirac) Z. Phys. {\bf 69}, 742. \\[2mm]
Schwinger, J. (1951). On gauge invariance and vacuum polarization,
Phys. Rev. {\bf 82}, 664–679. \\[2mm]
Sokolov, A. A. and Ternov I. M. (Editors), {\it The Synchrotron
Radiation}, (Nauka, Moscow, 1966). (in Russian).\\[2mm]
Yanovsky, V.,  Chvykov, V., Kalinchenko, G., Rousseau, P., Planchon, T., Matsuoka, T., Maksimchuk, A., Nees, J., Cheriaux, G., Mourou, G. and Krushelnick, K. (2008). Ultra-high intensity- 300-TW laser at 0.1 Hz repetition rate,
Optics Express, {\bf 16} Issue 3, 2109-2114. \\[2mm]
Wolkow, D. M. (1935). $\ddot{\rm U}$ber eine Klasse von L$\ddot{\rm o}$sungen der Diracschen Gleichung, Z. Physik, {\bf 94}, 250 - 260.\\[2mm]
Zel'dovich, Ya. B. (1975). Interaction of free electrons with electromagnetic radiation, Sov. Phys. Usp. {\bf 115}, No. 2, 161. \\[2mm]
Zhidkov, A., Koga, J., Sasaki, A. and  Uesaka, M. (2002).  Radiation damping effects on the interaction of ultraintense laser pulses with an overdense plasma, Phys. Rev. Lett. {\bf 88}, 185002. 
\end{document}